\documentclass{ws-procs9x6}

\usepackage{graphicx}
\newcommand{\be}{\begin{equation}}
\newcommand{\ee}{\end{equation}}
\newcommand{\bea}{\begin{eqnarray}}
\newcommand{\eea}{\end{eqnarray}}

\newcommand{\xbj}{x_{\!\scriptscriptstyle B}}

\newcommand{\bfk}{\mbox{\boldmath $k$}}
\newcommand{\bfq}{\mbox{\boldmath $q$}}
\newcommand{\bfP}{\mbox{\boldmath $P$}}

\newcommand{\pup}{p^\uparrow}

\newcommand{\bfp}{\mbox{\boldmath $p$}}

\newcommand{\bfS}{\mbox{\boldmath $S$}}

\def\lsim{\mathrel{\rlap{\lower4pt\hbox{\hskip1pt$\sim$}}\raise1pt\hbox{$<$}}}
\def\gsim{\mathrel{\rlap{\lower4pt\hbox{\hskip1pt$\sim$}}\raise1pt\hbox{$>$}}}
\def\nostrocostruttino#1\over#2{\mathrel{\mathop{\kern 0pt \rlap
{\hbox{$#1$}}} \hbox{\kern-.135em $#2$}}}

\newcommand{\PL}[1]{{\it Phys.\ Lett.}\ {\bf #1}}
\newcommand{\PR}[1]{{\it Phys.\ Rev.}\ {\bf #1}}
\newcommand{\PRL}[1]{{\it Phys.\ Rev.\ Lett.}\ {\bf #1}}

\def\kt{k_\perp}

\def\ptq{p_\perp}

\begin{document}

\title{Sivers function: SIDIS data, fits and predictions\footnote{\uppercase{T}alk presented by \uppercase{A}. \uppercase{P}rokudin}} 
\author{M.~Anselmino$^1$, M.~Boglione$^1$, U.~D'Alesio$^2$, A. Kotzinian$^3$, 
F.~Murgia$^2$, A. Prokudin$^1$}
\address{$^1$Dipartimento di Fisica Teorica, Universit\`a di Torino and \\
          INFN, Sezione di Torino, Via P. Giuria 1, I-10125 Torino, Italy\\
$^2$INFN, Sezione di Cagliari and Dipartimento di Fisica,  
Universit\`a di Cagliari,\\
C.P. 170, I-09042 Monserrato (CA), Italy\\ 
$^3$Dipartimento di Fisica Generale, Universit\`a di Torino and \\
          INFN, Sezione di Torino, Via P. Giuria 1, I-10125 Torino, Italy}
\maketitle

\abstracts{
\noindent
The most recent data on the weighted transverse single spin asymmetry 
$A_{UT}^{\sin(\phi_h-\phi_S)}$ from HERMES and COMPASS 
collaborations are analysed within LO parton model; all transverse motions are taken 
into account. 
Extraction of the Sivers function for $u$ and $d$ quarks is performed. 
Based on the  extracted Sivers functions, predictions for $A_{UT}^{\sin(\phi_h-\phi_S)}$ 
asymmetries at JLab are given; suggestions for further measurements at 
COMPASS, with a transversely polarized hydrogen target and selecting 
favourable kinematical ranges, are discussed. Predictions are also presented 
for Single Spin Asymmetries (SSA) in Drell-Yan processes at RHIC and GSI.  
}
\vspace{0.6cm}

\section{ Introduction }
In recent papers\cite{ourpaper}$^,$\cite{ourpaper1} we have discussed the role of intrinsic 
motions in inclusive and Semi-Inclusive Deep Inelastic Scattering (SIDIS) 
processes, both in unpolarized and polarized $\ell \, p \to \ell \, h \, X$ 
reactions. The LO QCD parton model computations have been compared with data on Cahn effect\cite{cahn}; this allows an estimate of the average values of the 
transverse momenta of quarks inside a proton, $\bfk_\perp$, and of final 
hadrons inside the fragmenting quark jet, $\bfp_\perp$, with the best fit 
results:
$\langle\kt^2\rangle   = 0.25  \;({\rm GeV}/c)^2$,
$ \langle\ptq^2\rangle  = 0.20 \;({\rm GeV}/c)^2$.

More detail, both about the kinematical configurations and conventions~\cite{trento} and the fitting procedure can be found in Ref. [\refcite{ourpaper}]. 

Equipped with such estimates, we have studied\cite{ourpaper}$^,$\cite{ourpaper1} the 
transverse single spin asymmetries $A_{UT}^{\sin(\phi_{\pi}-\phi_S)}$ observed 
by HERMES collaboration\cite{hermnew} and COMPASS collaboration\cite{compnew}; that allowed extraction 
of the Sivers function\cite{siv}
\be 
\Delta^N \! f_ {q/\pup}(x,k_\perp) = - \frac{2\,k_\perp}{m_p} \> 
f_{1T}^{\perp q}(x, k_\perp) \>, \label{rel} 
\ee
defined by
\bea
f_ {q/\pup} (x,\bfk_\perp) &=& f_ {q/p} (x,\kt) +
\frac{1}{2} \, \Delta^N \! f_ {q/\pup}(x,\kt)  \;
{\bfS} \cdot (\hat {\bfP}  \times
\hat{\bfk}_\perp) \;,\label{sivnoi} 
\eea
where $f_ {q/p}(x,\kt)$ is the unpolarized $x$ and $\kt$ dependent 
parton distribution ($\kt = |\bfk_\perp|$); $m_p$, $\bfP$ and $\bfS$ are 
respectively the proton mass, momentum and transverse polarization vector
($\hat{\bfP}$ and $\hat{\bfk}_\perp$ denote unit vectors). 

 We consider here these 
whole new sets of HERMES\cite{hermnew} and COMPASS\cite{compnew} data and perform a novel fit\cite{ourpaper1} of the 
Sivers functions. It turns out that the data well constrain  the 
parameters, thus offering the first direct significant estimate of the 
Sivers functions -- for $u$ and $d$ quarks -- active in SIDIS processes.
The sea quark contributions are found to be negligible, at least in the 
kinematical region of the available data.  
Finally, we exploit the QCD prediction\cite{col} 
$f_{1T}^{\perp q}(x, k_\perp)|_{\rm{D-Y}} = -f_{1T}^{\perp q}(x, k_\perp)|_{\rm{DIS}}$
and compute a single spin asymmetry, which can only originate from the Sivers
mechanism\cite{noi2}, for Drell-Yan processes at RHIC and GSI. The issue of QCD factorization of 
SIDIS and Drell-Yan processes was studied in Ref.[\refcite{Ji:2004wu}].

\section{ Extracting the Sivers functions }
Following Ref.[\refcite{ourpaper}], the inclusive ($\ell \, p \to \ell \, X $)
unpolarized DIS cross section in non collinear LO parton model is given by 
\be
\frac{d^2 \sigma ^{\ell p\to \ell X}}{d \xbj \, dQ^2} = 
\sum_q \int {d^2 \bfk _\perp}\; f_q(x,\kt) \; 
\frac{d\hat\sigma ^{\ell q\to \ell q}}{dQ^2} \;
J(\xbj, Q^2, \kt) \>,
\label{dsigkt}
\ee
and the semi-inclusive one ($\ell \, p \to \ell \, h \, X $) by
\be
\frac{d^5\sigma^{\ell p \to \ell h X }}{d\xbj \, dQ^2 \, dz_h \, d^2 \bfP _T} =
\sum_q \int {d^2 \bfk _\perp}\; f_q(x,\kt) \;
\frac{d \hat\sigma ^{\ell q\to \ell q}}{dQ^2} \;
J\; \frac{z}{z_h} \; D_q^h(z,p _\perp) \>,
\label{sidis-Xsec-final} 
\ee 
where  
\be
J = \frac{\hat s^2}{\xbj^2 s^2} \; \frac{\xbj}{x}
\left( 1 + \frac{\xbj^2}{x^2}\frac{\kt^2}{Q^2} \right)^{\!\!-1}  
\label{fcnj}, \frac{d \hat\sigma^{\ell q\to \ell q}}{d Q^2} = e_q^2 \, 
\frac{2\pi \alpha^2}{\hat s^2}\,
\frac{\hat s^2+\hat u^2}{Q^4}\;\cdot
\ee

$Q^2$, $\xbj$ and $y = Q^2/(\xbj\,s)$ are the usual leptonic DIS variables 
and $z_h, \bfP_T$ the usual hadronic SIDIS ones, in the $\gamma^*$--$p$ c.m. 
frame; $x$ and $z$ are light-cone momentum fractions, with (see Ref.[\refcite{ourpaper}] for exact relationships and further detail):
$
x = \xbj  + \mathcal{ O} \left(\frac{\kt^2}{Q^2} \right)$, 
$z = z_h   + \mathcal{ O} \left(\frac{\kt^2}{Q^2} \right)$,
$\bfp_\perp =  \bfP_T - z_h \, \bfk _\perp + 
\mathcal{ O}\left(\frac{\kt^2}{Q^2} \right)
$.

The $\sin(\phi_h-\phi_S)$ weighted transverse single spin asymmetry,
measured by HERMES and COMPASS, which singles out the contribution of the 
Sivers function (\ref{rel}), is given by:
\bea
A^{\sin (\phi_h-\phi_S)}_{UT} &=& \Big [ \displaystyle  \sum_q \! \int \!\! 
{d\phi_S \, d\phi_h \, d^2 \bfk _\perp}\;
\Delta ^N \! f_{q/\pup} (x,\kt) \sin (\varphi -\phi_S) \cdot \; 
  \label{hermesut} \nonumber \\
&&\frac{d \hat\sigma ^{\ell q\to \ell q}}{dQ^2} J\; \frac{z}{z_h} \; D_q^h(z,p _\perp) \sin (\phi_h -\phi_S)
\Big ]/ \\
&&\Big [ \displaystyle \sum_q \! \int \!\! {d\phi_S \, d\phi_h}\; \cdot 
\displaystyle {d^2 \bfk _\perp}\; 
f_{q/p}(x,\kt) \; \frac{d \hat\sigma ^{\ell q\to \ell q}}{dQ^2} \;
J\; \frac{z}{z_h} \; D_q^h(z,p _\perp) \Big ] \, , \nonumber
\eea
where $\varphi$ is the azimuthal angle of the quark transverse momentum, 
$\phi_h$ and $\phi_S$ are the azimuthal angles of produced hadron and polarization vector correspondingly.
We shall use Eq. (\ref{hermesut}), in which we insert a parameterization for 
the Sivers functions, to fit the experimental data.

The $k _\perp$ integrated parton distribution and fragmentation functions $f_q(x)$ and
$D_q^h(z)$ are taken from the literature, at the appropriate $Q^2$ values 
of the experimental data \cite{mrst01,kre}. 

We parameterize, for each light quark flavour $q=u,d$, the 
Sivers function in the following factorized form:
\be
\Delta^N \! f_ {q/\pup}(x,\kt) = 2 \, \mathcal{ N}_q(x) \, h(\kt) \, 
f_ {q/p} (x,\kt)\; , \label{sivfac}
\ee
where
\be
\mathcal{ N}_q(x) =  N_q \, x^{a_q}(1-x)^{b_q} \,
\frac{(a_q+b_q)^{(a_q+b_q)}}{a_q^{a_q} b_q^{b_q}}\; , 
h(\kt) = \frac{2\kt \, M_0}{\kt^2+ M_0^2}\; \cdot
\label{siverskt}
\ee
$N_q$, $a_q$, $b_q$ and $M_0$ (GeV/$c$) are free parameters. 
$f_ {q/p} (x,\kt)$ is the unpolarized distribution function. Since $h(\kt) \le 1$ and since we allow the constant 
parameter $N_q$ to vary only inside the range $[-1,1]$ so that 
$|\mathcal{ N}_q(x)| \le 1$ for any $x$, the positivity bound for the Sivers 
function is automatically fulfilled:
\be
\frac{|\Delta^N\!f_ {q/\pup}(x,\kt)|}{2 f_ {q/p} (x,\kt)}\le 1\; . \label{pos}
\ee

We neglect the 
contributions of sea quark functions and consider only the contributions of 
$\Delta^N \! f_{u/\pup}$ and $\Delta^N \! f_{d/\pup}$, for a total of 7 free 
parameters:
\be
N_u \quad a_u \quad b_u \quad N_d \quad a_d \quad b_d \quad M_0.
\label{par}
\ee 

\begin{figure}[h]
\includegraphics[height=0.18\textheight,width=0.5\textwidth,angle=-90,bb= 20 30 570 450]{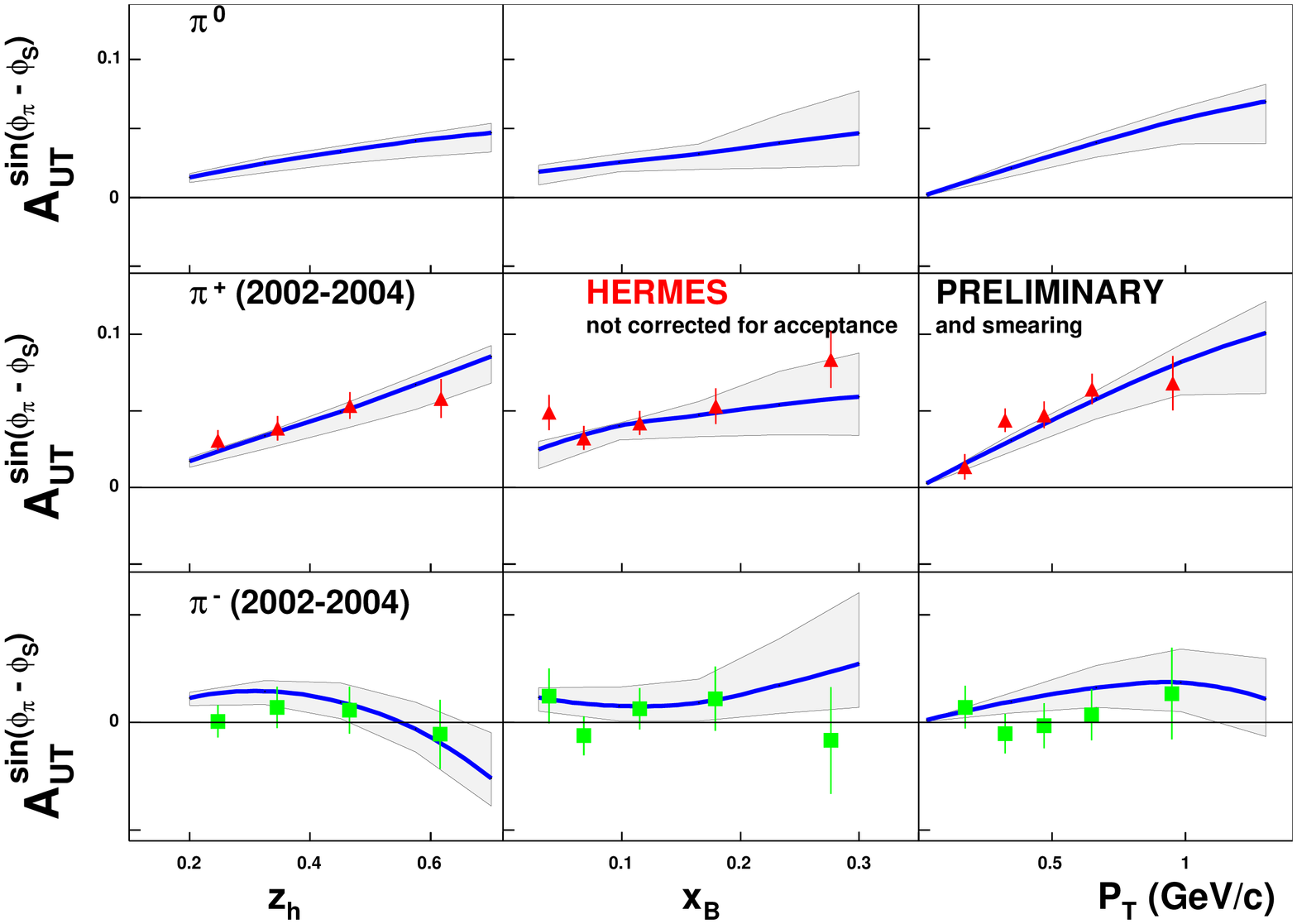}
\hskip 2.cm\includegraphics[height=0.18\textheight,width=0.5\textwidth,angle=-90,bb= 20 30 570 450]{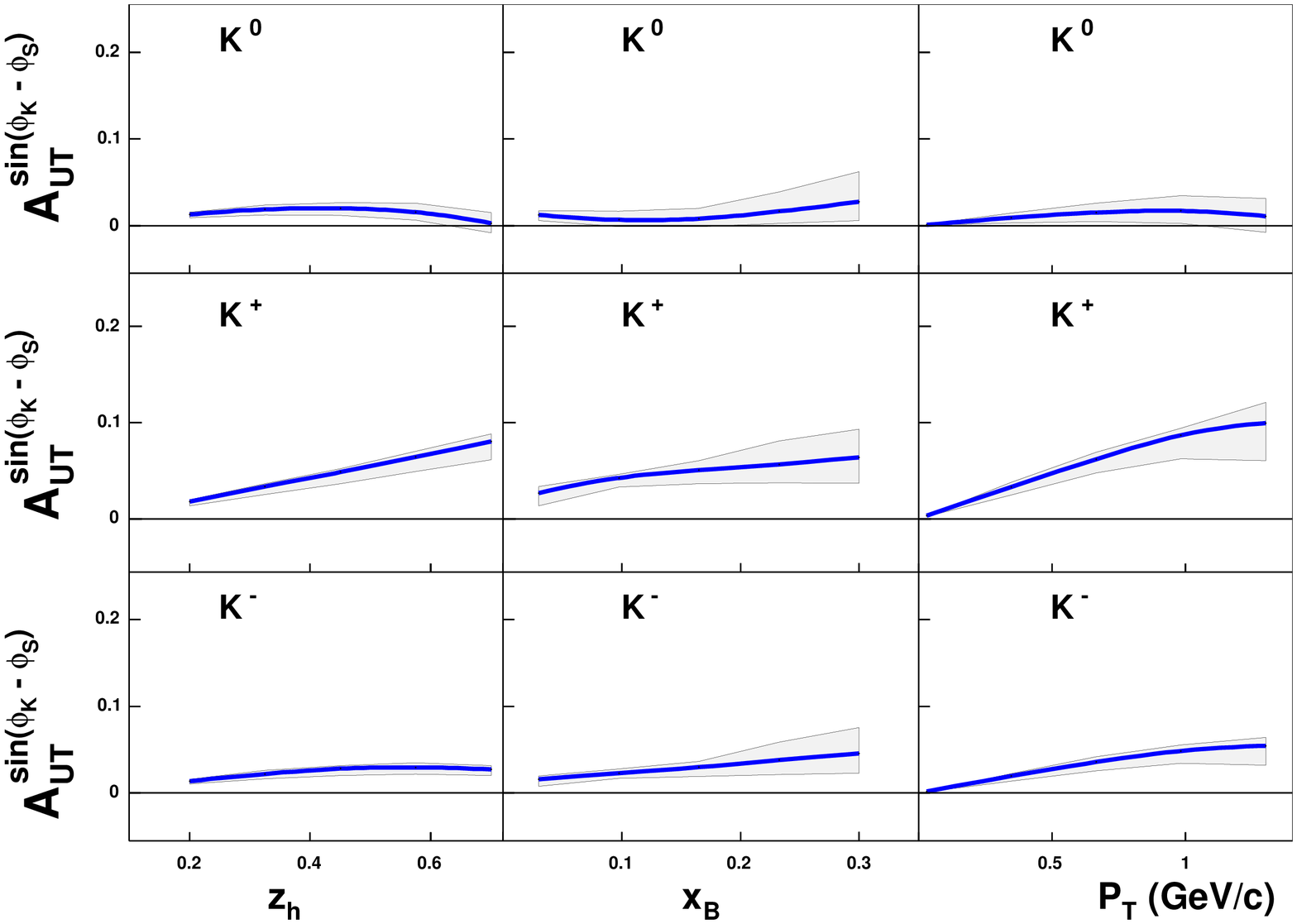}
\caption{HERMES data on $A_{UT}^{\sin(\phi_\pi-\phi_S)}$ for scattering
off a transversely polarized proton target and charged pion production. The 
curves are the results of our fit. The shaded area spans a region 
corresponding to one-sigma deviation at 90\% CL. 
Predictions for $\pi^0$ (upper-left panel) and kaon (right panels) asymmetries are 
also shown.\label{fig:authermes}}
\end{figure}

\begin{figure}[h]
\includegraphics[height=0.35\textheight,width=0.8\textwidth,bb= 20 30 570 450]
{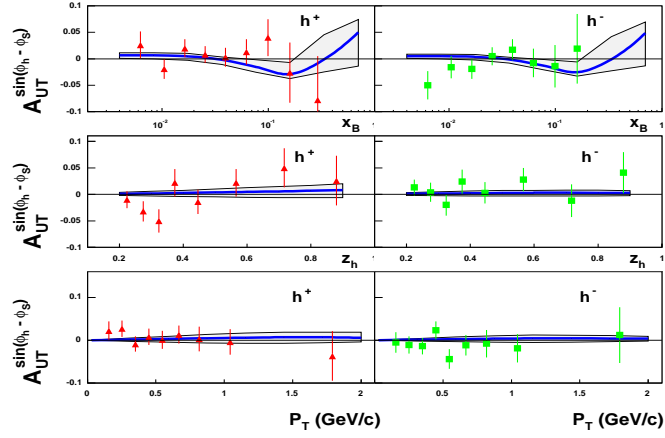}
\caption{COMPASS data on $A_{UT}^{\sin(\phi_h-\phi_S)}$ for scattering 
off a transversely polarized deuteron target and the production of positively
($h^+$) and negatively ($h^-$) charged hadrons. The curves are
the results of our fit. The shaded area spans a region corresponding to 
one-sigma deviation at 90\% CL.}
\label{fig:autcompass}
\end{figure}

The results of our fits are shown in Figs. \ref{fig:authermes} and
\ref{fig:autcompass}. 

In Fig.\ref{fig:authermes} we also show predictions, obtained using the extracted Sivers 
functions (see Table \ref{fitpar}), for $\pi^0$ and $K$ production; data on these asymmetries might 
be available soon from HERMES collaboration.

\begin{table}[ph]
\tbl{Best fit values of the parameters of the Sivers functions.}
{\footnotesize
\begin{tabular}{@{}rrrr@{}}
\hline
$N_{u}$  = & $0.32  \pm  0.11$ & $N_{d}$ = & $-1.00  \pm  0.12$ \\
$a_{u}$  = & $0.29  \pm  0.35$ & $a_{d}$ = & $ 1.16  \pm  0.47$ \\
$b_{u}$  = & $0.53  \pm  3.58$ & $b_{d}$ = & $ 3.77  \pm  2.59$ \\
\hline
$M_0^2$ = & $0.32 \pm 0.25 \; {\rm (GeV}/c)^2$ & $\chi^2/{d.o.f.}$ = 
& $1.06$ \\
\hline
\end{tabular}\label{fitpar} }
\vspace*{-13pt}
\end{table}

\section{ $A_{UT}^{\sin(\phi_h-\phi_S)}$ at COMPASS with polarized hydrogen 
target} 

By inspection of Eq. (\ref{hermesut}) it is easy to understand our numerical 
results for the $u$ and $d$ Sivers functions. In fact one can see that for 
scattering off a hydrogen target (HERMES), one has 
\be
\left( A^{\sin (\phi_h-\phi_S)}_{UT}\right)_{\rm hydrogen} \sim
4 \, \Delta^N \! f_{u/\pup} \, D_u^h + \Delta^N \! f_{d/\pup} \, D_d^h \>,
\label{hydt}
\ee
while, for scattering off a deuterium target (COMPASS),  
\be
\left( A^{\sin (\phi_h-\phi_S)}_{UT}\right)_{\rm deuterium} \sim
\left( \Delta^N \! f_{u/\pup} + \Delta^N \! f_{d/\pup} \right)
\left( 4 \, D_u^h + D_d^h \right) \>.
\ee
Opposite $u$ and $d$ Sivers contributions suppress COMPASS asymmetries for
any hadron $h$. 

\begin{figure}[h]
\includegraphics[height=0.26\textheight,width=0.5\textwidth,bb= 20 30 570 450]
{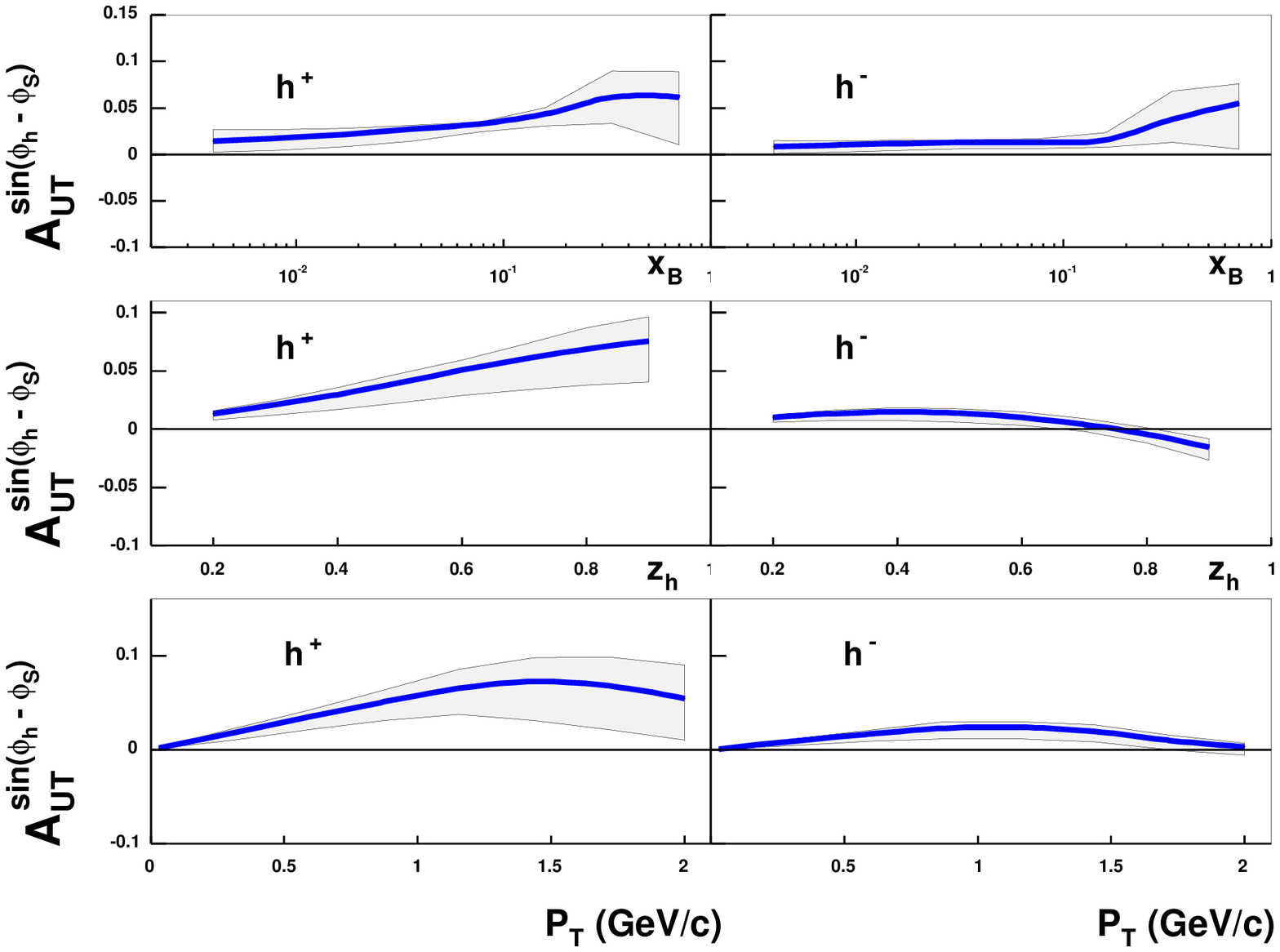}~\hfill
\includegraphics[height=0.26\textheight,width=0.5\textwidth,bb= 20 30 570 450]
{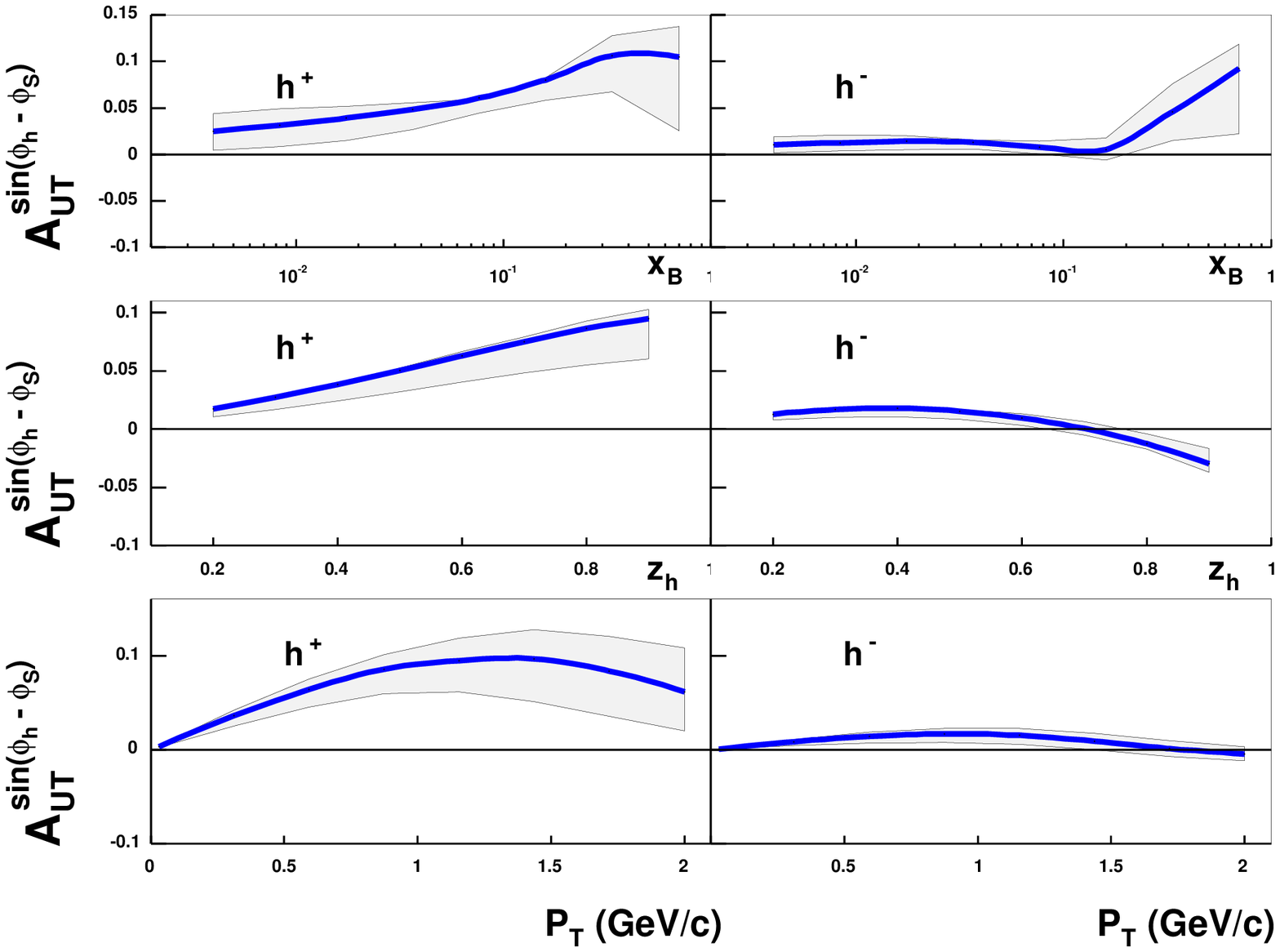}
\caption{Predictions for $A_{UT}^{\sin(\phi_h-\phi_S)}$ at COMPASS for scattering off 
a transversely polarized proton target and the production of positively
($h^+$) and negatively ($h^-$) charged hadrons. The plots in the left panel
have been obtained by performing the integrations over the unobserved 
variables according to the standard COMPASS kinematical cuts;
results with suggested new cuts, Eq. (\ref{new cuts}), are presented in the 
right panel.}
\label{fig:autcompass_proton}
\end{figure}

However, the COMPASS collaboration will soon be taking data with a transversely
polarized hydrogen target. 
Adopting the same experimental cuts which were used for the deuterium target\cite{ourpaper} the asymmetry is found to be around 5\% (see Fig. \ref{fig:autcompass_proton}).
These expected values can be further increased by properly selecting the 
experimental data. For example, selecting events with
\be
0.4 \le z_h \le 1 \quad\quad
0.2 \le P_T \le 1 \; {\rm GeV}/c \quad\quad
0.02 \le \xbj \le 1 \;,
\label{new cuts}
\ee 
yields the predictions shown in the right panel of  
Fig. \ref{fig:autcompass_proton}. The asymmetry for positively charged hadrons
becomes larger, and one 
expects a clear observation of a sizeable azimuthal asymmetry also for the 
COMPASS experiment.

\section{$A_{UT}^{\sin(\phi_h-\phi_S)}$ at JLab with polarized hydrogen 
target}

\begin{figure}[h]
\includegraphics[width=0.5\textwidth,bb= 20 30 570 450]
{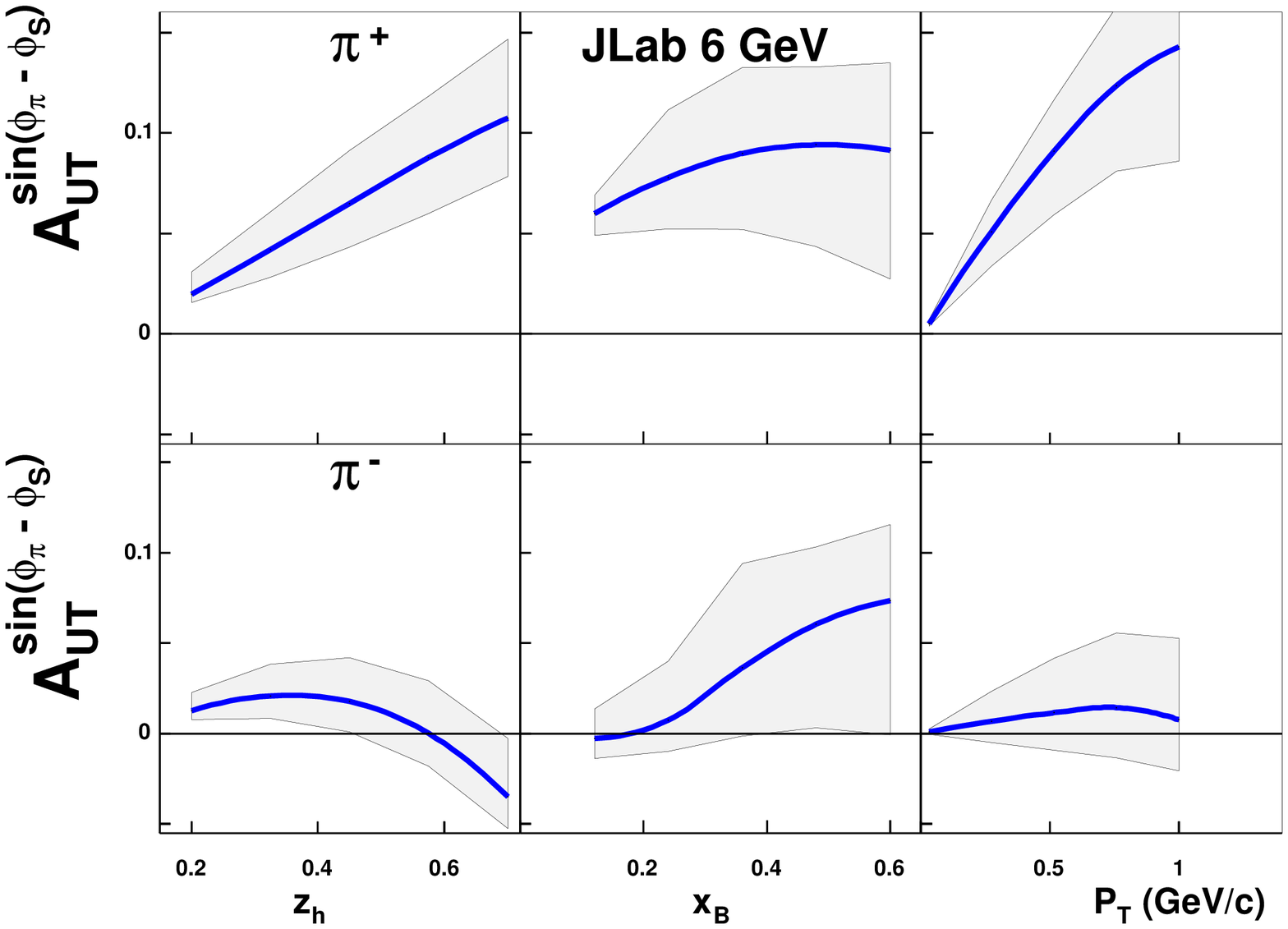}~\hfill
\includegraphics[width=0.5\textwidth,bb= 20 30 570 450]
{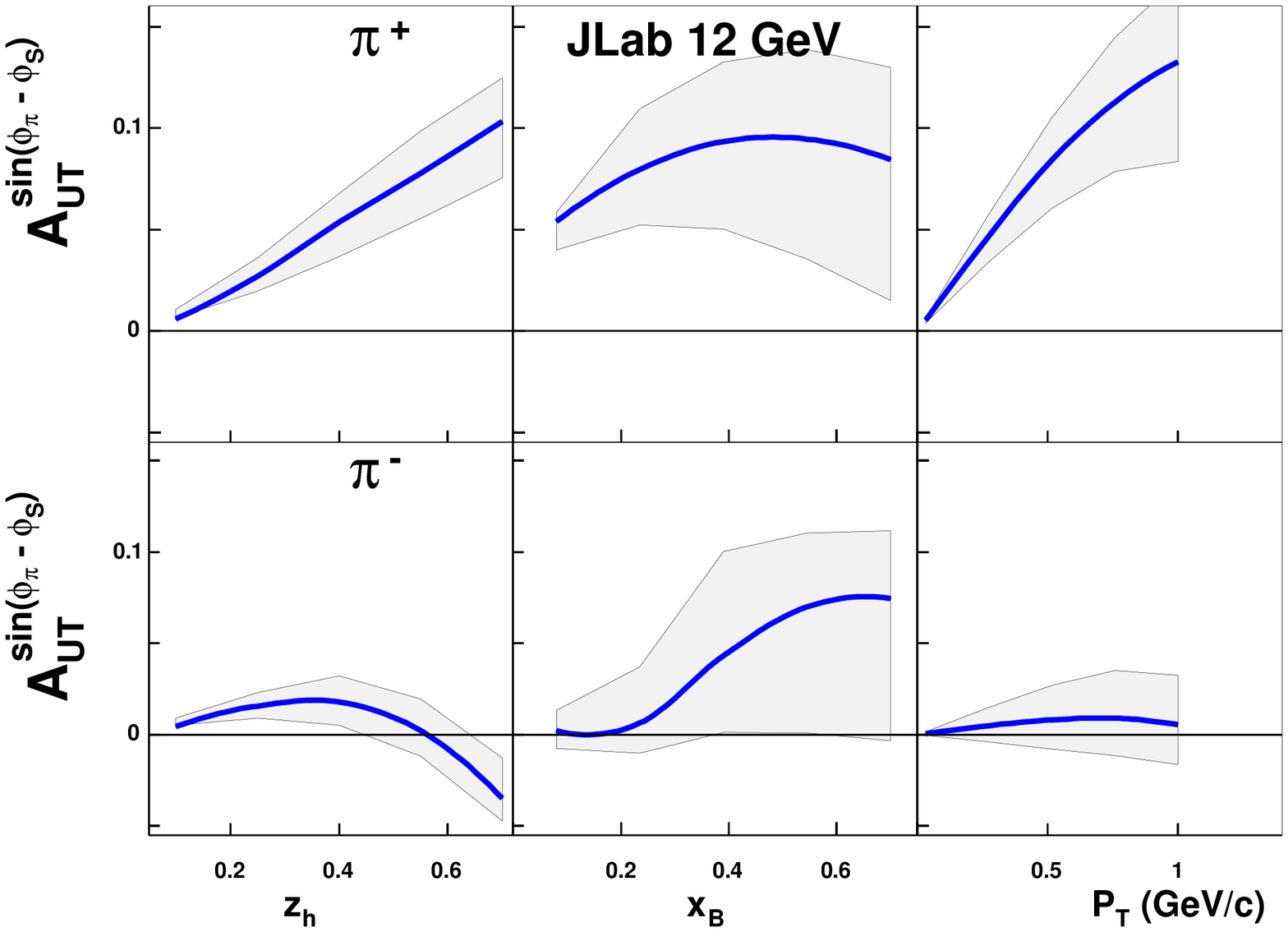}
\caption{Predictions for $A_{UT}^{\sin(\phi_{\pi}-\phi_S)}$ at JLab for the production 
of $\pi^+$ and $\pi^-$ from scattering off a transversely polarized proton 
target.}
\label{fig:autjlab}
\end{figure}

Also JLab experiments are supposed to measure the SIDIS azimuthal asymmetry 
for the production of pions on a transversely polarised hydrogen target, at 
incident beam energies of 6 and 12 GeV. The kinematical region of this 
experiment is very interesting, as it will supply information on the 
behaviour of the Sivers functions in the large-$\xbj$ domain, up to 
$\xbj \simeq 0.6$. 
Imposing the experimental cuts of JLab we obtain the predictions shown in 
Fig. \ref{fig:autjlab}. A large and healthy azimuthal asimmetry for 
$\pi^+$ production should be observed. Similar results have been obtained 
also in an approach based on a Monte Carlo event generator \cite{aram}.  

\section{ Transverse SSA in Drell-Yan processes}

Let us now consider the transverse single spin asymmetry, 
\be
A_N = \frac{d\sigma^\uparrow - d\sigma^\downarrow}
           {d\sigma^\uparrow + d\sigma^\downarrow} \> , \label{asy}
\ee
for Drell-Yan processes, $\pup \, p \to \ell^+ \, \ell^- \, X$, 
$\pup \, \bar p \to \ell^+ \, \ell^- \, X$ and 
$\bar p^{\,\uparrow} \, p \to \ell^+ \, \ell^- \, X$, where $d\sigma$ 
stands for ${d^4\sigma}/{dy \, dM^2 \, d^2\bfq_T}$
and $y$, $M^2$ and $\bfq_T$ are respectively the rapidity, the squared 
invariant mass and the transverse momentum of the lepton pair in the initial
nucleon c.m. system. 

In such a case the SSA (\ref{asy}) can only originate 
from the Sivers function and is given (selecting the region with 
$q_T^2 \ll M^2, \> q_T \simeq k_\perp$) by \cite{noi2} 
\bea
\nonumber
A_N & = &\Big [ \sum_q e_q^2 \int d^2\bfk_{\perp q} \, d^2\bfk_{\perp \bar q} \>
\delta^2(\bfk_{\perp q} + \bfk_{\perp \bar q} - \bfq_T)  \>\Delta^N \! f_{q/\pup}(x_q, \bfk_{\perp q})
\cdot \nonumber\\
&&f_{\bar q/p}(x_{\bar q}, \bfk_{\perp \bar q}) \Big ] \, / \, \Big [ 2 \sum_q e_q^2 \int d^2\bfk_{\perp q} \, d^2\bfk_{\perp \bar q} \>\delta^2(\bfk_{\perp q} + \bfk_{\perp \bar q} - \bfq_T) 
 \cdot \nonumber \\
&&f_{q/p}(x_q, \bfk_{\perp q}) \>
f_{\bar q/p}(x_{\bar q}, \bfk_{\perp \bar q})\Big ]\, , \label{ann}
\eea
where $q = u, \bar u, d, \bar d, s, \bar s$ and
$x_q = \frac{M}{\sqrt s} \, e^y$, 
$x_{\bar q} = \frac{M}{\sqrt s} \, e^{-y}$.
Eq. (\ref{ann}) explicitely refers to $\pup\,p$ processes, with obvious 
modifications for $\pup \, \bar p$ and $\bar p^{\, \uparrow} \, p$ ones.  

\begin{figure}[h]
\includegraphics[width=0.25\textwidth,angle=90,bb= 540 50 200 770]
{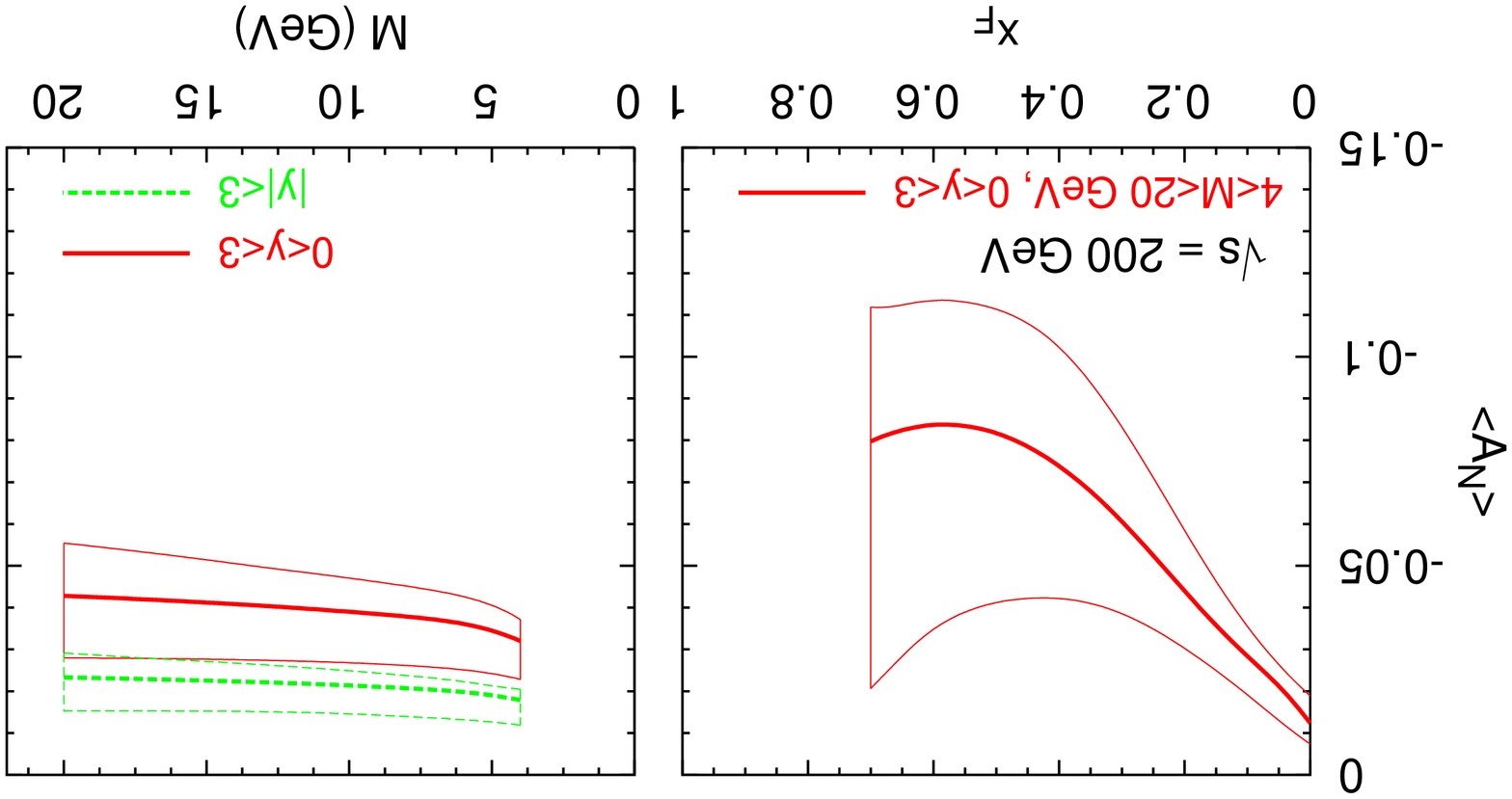}\hskip 6.cm
\includegraphics[width=0.25\textwidth,angle=90,bb= 540 50 200 770]
{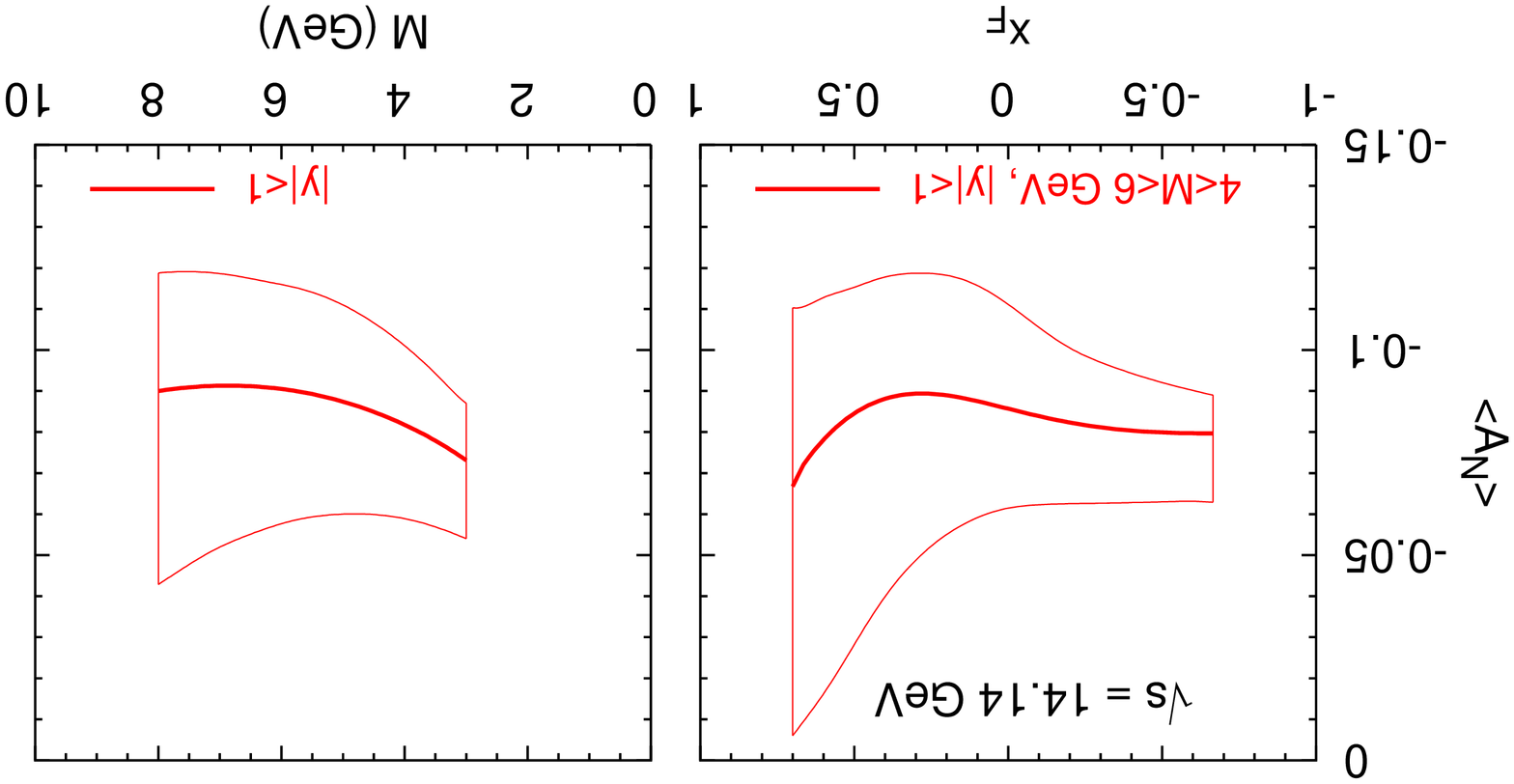}
\caption{Predictions for single spin asymmetries in Drell-Yan, $\pup \, p \to \ell^+ \, \ell^- \, X$, processes at RHIC (left panel) and GSI (right panel), according to Eq. (\ref{ann}) of the 
text. }
\label{fig:anrhic}
\end{figure}

Inserting into Eq. (\ref{ann}) the Sivers functions extracted from our fit 
to SIDIS data and {\it reversed in sign}\cite{col}, 
we obtain the predictions (shown in Fig. \ref{fig:anrhic}) for RHIC (left panel) and PAX (right panel) experiment \cite{pax} planned at 
the proposed asymmetric $p \, \bar p$ collider at GSI.       

\section{ Comments and conclusions}

The  Sivers functions $\Delta^N \! f_ {u/\pup}(x,k_\perp)$ and $\Delta^N \! f_ {d/\pup}(x,k_\perp)$
have been extracted using recent HERMES \cite{hermnew} and COMPASS \cite{compnew} collaborations
data on $A_{UT}^{\sin(\phi_h-\phi_S)}$.

A sizeable $h^+$ asymmetry should be measured by COMPASS 
collaboration once they switch, as planned, to a transversely polarized 
hydrogen target.  

Large values of $A_{UT}^{\sin(\phi_h-\phi_S)}$ are expected at JLab, both 
in the 6 and 12 GeV operational modes, for $\pi^+$ inclusive production.

We have then used basic QCD relations and computed the single 
spin asymmetries in Drell-Yan processes. We have used the same Sivers functions 
as extracted from SIDIS data, with opposite signs. The predicted $A_N$ could 
be measured at RHIC in $p\,p$ collisions and, in the long range, at the 
proposed PAX experiment at GSI\cite{pax}, in $p\,\bar p$ interactions. 
It would provide a clear and stringent test of basic QCD properties. 




\end{document}